\documentclass[manuscript]{acmart}
\usepackage{multirow}
\usepackage{multicol}
\usepackage{array}    
\usepackage{framed}
\usepackage{hyperref}
\usepackage[capitalise,noabbrev]{cleveref}
\usepackage{listings}
\usepackage{xspace}
\usepackage{csquotes}
\usepackage{dblfloatfix}
\usepackage{placeins}
\usepackage{soul}
\usepackage{microtype}
\usepackage{subcaption}
\usepackage{etoolbox}
\usepackage{colortbl}
\usepackage{makecell}
\usepackage{pifont}
\usepackage{tabularx}
\usepackage{algorithm}
\usepackage{algpseudocode}
\usepackage{amsmath}
\usepackage{tikz}
\usepackage{wasysym}
\usepackage{tcolorbox}
\usepackage{booktabs}
\usepackage[table]{xcolor}

\newcolumntype{C}[1]{>{\centering\arraybackslash}m{#1}}
\newcolumntype{L}[1]{>{\raggedright\arraybackslash}m{#1}}

\definecolor{color-a}{RGB}{244, 241, 222}
\definecolor{color-b}{RGB}{129, 178, 154}
\definecolor{color-c}{RGB}{61, 64, 91}
\definecolor{color-d}{RGB}{242, 204, 143}
\definecolor{color-e}{RGB}{224, 122, 95}
\definecolor{color-f}{RGB}{201, 228, 202}
\definecolor{color-g}{RGB}{254, 217, 183}
\definecolor{gray}{gray}{0.9}
\definecolor{light-gray}{gray}{0.4}

\newcommand{\ourmethod}{\textsc{CoSA}\xspace}

\sethlcolor{cyan}

\renewcommand{\paragraph}[1]{\vspace{6pt}\noindent{\bf #1}\hspace{8pt}}

\newcommand\realnumberstyle[1]{}
\makeatletter
\newcommand{\linecolor}[3]{
    {\realnumberstyle{#3}}
    \begingroup
    \lst@basicstyle
    \ifnum\value{lstnumber}=#1
        \color{#2}
    \else
        \color{white}
    \fi
    \rlap{\hspace*{\lst@numbersep}
    \color@block{\linewidth}{\ht\strutbox}{\dp\strutbox}
    }
    \endgroup
}

\makeatletter
\newcommand{\storelinecolor}[2]{%
    \expandafter\gdef\csname linecolor@#1\endcsname{#2}%
}

\newcommand{\clearlinecolors}{%
    \@for\next:=1,2,3,4,5,6,7,8,9\do{%
        \expandafter\global\expandafter\let\csname linecolor@\next\endcsname\undefined
    }%
}

\newcommand{\multilinecolor}[1]{%
    {\realnumberstyle{#1}}%
    \begingroup
    \lst@basicstyle
    \ifcsname linecolor@\arabic{lstnumber}\endcsname
        \color{\csname linecolor@\arabic{lstnumber}\endcsname}%
    \else
        \color{white}%
    \fi
    \rlap{\hspace*{\lst@numbersep}%
    \color@block{\linewidth}{\ht\strutbox}{\dp\strutbox}%
    }%
    \endgroup
}

\makeatother

\newcommand{\linecolorrg}[3]{
    {\realnumberstyle{#3}}
    \begingroup
    \lst@basicstyle
    \ifnum\value{lstnumber}=#1
        \color{#2}
    \else
        \ifnum\value{lstnumber}>#1
            \ifnum\value{lstnumber}<\numexpr#1+\@firstofone#2\relax
                \color{\@secondoftwo#2}
            \else
                \color{white}
            \fi
        \else
            \color{white}
        \fi
    \fi
    \rlap{\hspace*{\lst@numbersep}
    \color@block{\linewidth}{\ht\strutbox}{\dp\strutbox}
    }
    \endgroup
}

\newcommand{\linecolorr}[4]{
    {\realnumberstyle{#4}}
    \begingroup
    \lst@basicstyle
    \ifnum\value{lstnumber}>=#1
        \ifnum\value{lstnumber}<=#2
            \color{#3}
        \else
            \color{red}
        \fi
    \else
        \color{white}
    \fi
    \rlap{\hspace*{\lst@numbersep}
    \color@block{\linewidth}{\ht\strutbox}{\dp\strutbox}
    }
    \endgroup
}

\newcommand{\linecolorrange}[4]{
    {\realnumberstyle{#4}}
    \begingroup
    \lst@basicstyle
    \ifnum\value{lstnumber}>=#1
        \ifnum\value{lstnumber}<=#2
            \color{#3}
        \else
            \color{white}
        \fi
    \else
        \color{white}
    \fi
    \rlap{\hspace*{\lst@numbersep}
    \color@block{\linewidth}{\ht\strutbox}{\dp\strutbox}
    }
    \endgroup
}

\newcommand{\colorlines}[1]{
    \renewcommand{\lst@DefEC}{%
        \lst@CCECUse \lst@ProcessLetter
        \global\let\lst@thestyle\relax
        \edef\tempa{#1}\edef\tempb{\thelstnumber}%
        \ifstrequal{\tempa}{\tempb}%
           {\global\def\lst@thestyle{\color{red}}}{}%
    }
    \expandafter\lst@AddToHook\expandafter{OutputOther}{\lst@thestyle}
}

\lstdefinelanguage{Diff}{
  language=Python,
  sensitive=true,
  morecomment=[f][\color{myred}]-,
  morecomment=[f][\color{mygreen}]+,
}

\AtBeginDocument{%
  }

\setcopyright{acmlicensed}
\copyrightyear{2018}
\acmYear{2018}
\acmDOI{XXXXXXX.XXXXXXX}

\acmJournal{JACM}
\acmVolume{37}
\acmNumber{4}
\acmArticle{111}
\acmMonth{8}

\begin{document}

\title{\ourmethod: Context-Aware Severity Assessment via Context Analysis with Large Language Models}
\author{Jinfeng Jiang}
\email{jfjiang@smu.edu.sg}
\affiliation{
  \institution{Singapore Management University}
  \city{Singapore}
  \country{Singapore}
}

\author{Yikun Li$^{*}$}
\email{yikunli@smu.edu.sg}
\affiliation{
    \institution{Singapore Management University}
    \city{Singapore}
    \country{Singapore}
}

\author{Chengran Yang}
\email{cryang@smu.edu.sg}
\affiliation{
  \institution{Singapore Management University}
  \city{Singapore}
  \country{Singapore}
}

\author{Ting Zhang}
\email{Ting.Zhang@monash.edu}
\orcid{https://orcid.org/0000-0002-6001-1372}
\affiliation{
  \institution{Monash University}
  \city{Melbourne}
  \country{Australia}
}

\author{Wen Bin Leow}
\email{leow_wen_bin@tech.gov.sg}
\affiliation{
  \institution{GovTech}
  \city{Singapore}
  \country{Singapore}
}

\author{Yide Yin}
\email{YIN_Yide@tech.gov.sg}
\affiliation{
  \institution{GovTech}
  \city{Singapore}
  \country{Singapore}
}

\author{Eng Lieh Ouh}
\email{elouh@smu.edu.sg}
\affiliation{
  \institution{Singapore Management University}
  \city{Singapore}
  \country{Singapore}
}

\author{Lwin Khin Shar}
\email{lkshar@smu.edu.sg}
\affiliation{
  \institution{Singapore Management University}
  \city{Singapore}
  \country{Singapore}
}

\author{David Lo}
\email{davidlo@smu.edu.sg}
\orcid{https://orcid.org/0000-0002-4367-7201}
\affiliation{
  \institution{Singapore Management University}
  \city{Singapore}
  \country{Singapore}
}

\thanks{$^{*}$ Yikun Li is the corresponding author (yikunli@smu.edu.sg).}

\renewcommand{\shortauthors}{Jiang et al.}

\begin{abstract}
Accurate vulnerability severity assessment is essential for prioritizing remediation, yet manually assessing Common Vulnerability Scoring System (CVSS) base metrics remains labor-intensive. Existing automated approaches often fail to capture the repository-level evidence required for assessing many CVSS base metrics. Such repository-aware assessment is challenging because relevant evidence is scattered across the entire repository under heavy noise.

To address these challenges, we present \ourmethod, a \textbf{Co}ntext-aware vulnerability \textbf{S}everity \textbf{A}ssessment approach that infers CVSS base metrics from repository artifacts. \ourmethod constructs a code property graph (CPG) and applies a two-stage repository-pruning strategy: lightweight static pruning to preserve structurally proximal context, followed by an agentic large language model (LLM)-guided pruning step to retain CVSS-relevant context while collecting supporting evidence. The LLM then consolidates the retrieved repository context into compact, CVSS metric-wise textual summaries, which are fed into a lightweight transformer predictor. We also construct a higher-quality repository-level dataset comprising 6,816 CVSS labeled instances spanning 90 Common Weakness Enumeration (CWE) types.

Experiments on real-world vulnerabilities show that \ourmethod consistently outperforms function-level and pure-LLM baselines. It improves prediction accuracy by 14.4\% and Macro-F1 by 15.3\% over the best-performing baseline, suggesting that explicit, metric-oriented repository context retrieval is crucial for practical and reliable automated severity assessment.
\end{abstract}



\maketitle

\section{Introduction}
\label{sec:intro}
Modern software systems are deeply intertwined. Cloud services, mobile applications, and open-source dependencies form supply-chain ecosystems in which a single flaw can lead to widespread compromise and operational disruption~\cite{liu_icse_npm_2022,shen_ssc_vuln_emse_2025}.
Meanwhile, public vulnerability reporting continues to accelerate: 39,962 Common Vulnerabilities and Exposures (CVEs) were published in 2024, increasing further to 48,185 in 2025~\cite{gamblin_cve_2025_review}. 
NIST reports that CVE submissions increased by 32\% in 2024, that the prior processing rate is no longer sufficient to keep up with incoming submissions, and that the National Vulnerability Database (NVD) backlog is still growing while exploring machine learning to automate parts of vulnerability processing~\cite{nist_nvd_general_update_2025}. 
On the other hand, automated repository and pipeline-level vulnerability detection tools often surface orders of magnitude more candidate issues than those ultimately curated and assigned CVE identifiers~\cite{huang_revisiting_2025}. In this setting, severity assessment becomes a practical bottleneck for vulnerability management, as it determines which issues should be fixed with higher priority, and how risks are communicated across technical and non-technical stakeholders.

In practice, severity assessment is primarily based on scorecards such as the Common Vulnerability Scoring System (CVSS)~\cite{mell_common_2006,FIRST_CVSS3.1_2019}. 
CVSS v3.x expresses the intrinsic exploitability and impact of a vulnerability through a series of base metrics and maps this set of features as a vector and normalizes them to a standardized severity band. It provides a standard for coordinating remediation between vendors, customers, and regulators. However, manually assigning accurate CVSS vectors is labor-intensive and requires expertise in both security and affected system. As the pace of disclosures increases~\cite{nist_nvd_general_update_2025,gamblin_cve_2025_review}, there is strong pressure to automate software vulnerability severity assessment.

Recent work therefore turns to automated vulnerability severity prediction. 
Based on the input signals they rely on, existing approaches can be roughly grouped into three categories: description-centric, hybrid, and code-centric approaches. 
Description-centric models take only natural-language vulnerability descriptions as input, and use classical machine learning or deep text encoders to predict CVSS severity or full base vectors from CVE descriptions~\cite{khazaei_automatic_2016,han_learning_2017,kudjo_improving_2019,elbaz_fighting_2020,shahid_cvss_bert_2021,shi_xlnet_cvss_2022,okutan_predicting_2022}. Hybrid approaches combine vulnerability descriptions with source code, prior CVSS vectors, and increasingly employ additional related web pages for prediction~\cite{dong_dekedver_2023,du_bimodal_mtl_2024,xue_prompt_tuning_sva_2025,gao_sva_icl_2025,shen2025vulstamp,Ye-KG4VA-2025}. 
However, the majority of these methods rely on the CVSS description, which means the severity can only be assigned after human analysis and reasoning, limiting the automation and efficiency of severity assessment. 
To address this limitation, researchers proposed several code-centric models that operate on commits, functions, or sliced vulnerable code~\cite{le_deepcva_2021,le_fine_grained_2022,li_commit-level_2023,nguyen_automated_2024,wen_evalsva_2024}, thus allowing for an earlier severity assessment that serves as a pre-processing step that accelerates following human intervention.

While prior works~\cite{pan_towards_practical_2024,li_commit-level_2023} have advanced automated severity assessment along multiple axes, accurate CVSS inference actually depends on a repository context that is large, scattered, and difficult to retrieve and organize. 
In practice, CVSS v3.x base metrics including Attack Vector (AV), Attack Complexity (AC), User Interaction (UI), Privileges Required (PR), Scope (S), and the impact metrics (C, I, A), which are rarely determined by the vulnerable function alone. 
Instead, they depend on how this function is reached through the program and how its effects propagate through subsequent operations. Consequently, the evidence needed for correct metric assignment is often distributed across multiple functions and call paths.
This observation motivates our work: repository-aware severity assessment. 
However, CVSS-labeled data is often released without the repository-level context needed to judge reachability and effect propagation.
Accordingly, repository-aware severity assessment faces four key challenges.

\textbf{(C0) Missing repository-level context in CVSS-labeled datasets.}
CVSS is usually attached to CVE/advisory records \cite{nist_nvd_cvss_metrics,ossf_osv_schema}, and downstream datasets often expose only commit, file, or function-level artifacts \cite{bhandari2021cvefixes,wen_evalsva_2024}. The repository-level context needed to decide reachability and effect propagation is rarely packaged, so reconstructing it from raw repositories is necessary.

\textbf{(C1) Context scale and dispersion.} CVSS-relevant evidence is frequently distributed across multi-hop call paths rather than localized within the vulnerable function, its file, or the patch hunk. As a result, narrow inputs used by existing works systematically omit the reachability and effect-propagation signals that govern base metric assignments.

\textbf{(C2) Relevance selection under heavy noise.} However, simple context expansion is expensive: real repositories exhibit branching call graphs, shared utility layers, and multiple usage scenarios. Much of this surrounding context is irrelevant to a specific vulnerability and can overwhelm downstream predictors. The key is to select a small, high-value subset of evidence that is most informative for CVSS.

\textbf{(C3) Metric-wise evidence organization for prediction.} Even after retrieving relevant context, the evidence should be consolidated into structured signals, so that predictions are grounded and stable rather than driven by unstructured long-context text.

To address these challenges, we propose \ourmethod, a \textbf{Co}ntext-aware vulnerability \textbf{S}everity \textbf{A}ssessment approach that infers CVSS v3.x base metrics from repository evidence. 
\ourmethod builds the code property graph (CPG), which is a representation of the repository. 
When a vulnerable function is detected, a two-stage pruning is performed to select key context. 
First, a lightweight static step restricts the search space to a structurally proximal neighborhood functions around the vulnerable function along call edges. 
This yields a compact candidate subgraph that still retains reachability and downstream effect paths that matter for CVSS decisions.
Second, an LLM-based agent operates over the pruned subgraph to perform CVSS-guided evidence gathering. The agent iteratively inspects a small set of candidates, extracts only the information that matters to metric decisions, and discards irrelevant branches. The gathered information is organized into metric-wise evidence summaries, which are then fed into a lightweight prediction model. In this way, we address (C0) by reconstructing repository-level evidence from raw repositories via a whole-repo CPG, enabling explicit reasoning about reachability and downstream effect propagation, (C1) by traversing a multi-hop repository structure, (C2) by pruning to retain only high-value evidence, and (C3) by producing structured summaries to compress the input length for efficient prediction.

Experimental results on real-world vulnerabilities demonstrate that our approach outperforms state of the art (SOTA)~\cite{wen_evalsva_2024} on CVSS v3.x base severity prediction. In particular, \ourmethod improves prediction accuracy by 14.4\% and macro-F1 by 15.3\% relative to the best-performing baseline, highlighting the importance of explicit, metric-oriented repository context retrieval for practical vulnerability management.

Our contributions are threefold:
\begin{itemize}
  \item We construct a higher-quality repository-level dataset including 6,816 CVSS v3.x labeled metric instances that span 90 Common Weakness Enumeration (CWE) types for repository-aware severity assessment by consolidating and reconciling existing CVSS-labeled vulnerability resources and filtering inconsistent or low-signal entries. 
  \item We propose \ourmethod, a scalable repository-evidence framework for CVSS v3.x base metric inference that explicitly models reachability and effect propagation through interprocedural structure while keeping the context budget controlled. By combining two-stage pruning with metric-wise evidence synthesis, \ourmethod turns dispersed repository signals into decision-relevant inputs, enabling more grounded and stable predictions than code-only approaches.
  \item We conduct extensive experiments and analysis on real-world vulnerabilities, showing consistent improvements over representative baselines on severity classification. Beyond aggregate gains, we provide diagnostic evidence that the retrieved and organized context is the primary driver of performance, offering actionable guidance for practical severity assessment systems.
\end{itemize}

\section{Background}

\subsection{Software Vulnerability Severity Assessment}

Many vulnerability severity indicators and prioritization frameworks are used in practice, including the Exploit Prediction Scoring System (EPSS)~\cite{jacobs_exploit_2019}, the Stakeholder-Specific Vulnerability Categorization (SSVC)~\cite{Spring2021SSVC}, the Common Weakness Scoring System (CWSS)~\cite{MITRE_CWSS}, and the OWASP Risk Rating~\cite{OWASP_RiskRating}, among them, the Common Vulnerability Scoring System (CVSS)~\cite{mell_common_2006} remains one of the most widely adopted standards for communicating the severity of software vulnerabilities~\cite{pan_towards_practical_2024,xue_prompt_tuning_sva_2025}. CVSS offers a metricized and reproducible procedure to quantify the intrinsic impact and exploitability of a flaw and to adapt that estimate over time and across environments. 
There are three major versions of CVSS in use today: v2 (2007)~\cite{FIRST_CVSS2.0_2007}, v3.x (2015–2019)~\cite{FIRST_CVSS3.0_2015, FIRST_CVSS3.1_2019}, and the recently released v4.0 (2024)~\cite{FIRST_CVSS4.0_2024}. Our work focuses on CVSS v3.x (including v3.0 and v3.1), which is currently the most prevalent version in vulnerability databases and industry practice~\cite{nvd_retire_cvss_v2_2022,gao_sva_icl_2025}.

\subsection{CVSS v3.x Base Metrics}

\begin{table}[h]
\centering
\footnotesize
\setlength{\tabcolsep}{3pt}
\renewcommand{\arraystretch}{1.05}
\caption{CVSS v3.x Base metrics, Allowed Values and Brief Meaning}
\label{tab:cvss31-base}
\begin{tabular}{@{}C{1.8cm} C{2.6cm} C{0.9cm} C{3.1cm} L{4.8cm}@{}}
\toprule
\multicolumn{1}{c}{\textbf{Group}} &
\multicolumn{1}{c}{\textbf{Metric}} &
\multicolumn{1}{c}{\textbf{Abbr.}} &
\multicolumn{1}{c}{\textbf{Allowed values}} &
\multicolumn{1}{c}{\textbf{Brief meaning}} \\
\midrule

\multirow{7}{*}{Exploitability} &
Attack Vector & AV &
N (Network), A (Adjacent), L (Local), P (Physical) &
How remotely an attacker can reach the vulnerable component. \\
\cline{2-5}
& Attack Complexity & AC &
L (Low), H (High) &
Conditions beyond the attacker's control that affect exploit success. \\
\cline{2-5}
& Privileges Required & PR &
N (None), L (Low), H~(High) &
Level of privileges needed prior to successful exploitation. \\
\cline{2-5}
& User Interaction & UI &
N (None), R (Required) &
Whether exploitation requires a separate user to take an action. \\
\midrule

\multirow{1}{*}{Scope} &
Scope & S &
U (Unchanged), C~(Changed) &
Whether exploitation can impact resources beyond the vulnerable component’s authorization boundary. \\
\midrule

\multirow{5}{*}{Impact} &
Confidentiality Impact & C &
N (None), L (Low), H~(High) &
The impact to confidentiality of a successfully exploited vulnerability. \\
\cline{2-5}
& Integrity Impact & I &
N (None), L (Low), H~(High) &
The impact to integrity of a successfully exploited vulnerability. \\
\cline{2-5}
& Availability Impact & A &
N (None), L (Low), H~(High) &
The impact to availability of a successfully exploited vulnerability. \\
\bottomrule
\end{tabular}
\end{table}

In CVSS version 3.x, the base metrics capture properties of a vulnerability that are intended to be stable across time and deployments. The Temporal metrics allow adjustment for evolving conditions such as exploit code maturity, while the Environmental metrics adapt the score to local conditions such as compensating controls or asset importance. Within the base group, CVSS~v3.x distinguishes Exploitability metrics, which describe the conditions and requirements for successfully exploiting a vulnerability, from Impact metrics, which describe the potential loss of confidentiality, integrity, or availability once exploitation occurs. It also includes Scope, which indicates whether successful exploitation can affect resources beyond the vulnerable component's own authorization boundary. Further descriptions of these metrics are listed in Table~\ref{tab:cvss31-base}.
The resulting base score is calculated by the values of these metrics, and mapped to standard base severity bands, normalizing indicators across different domains to serve as a unified reference for triage prioritization and cross-stakeholder communication.

CVSS v3.x computes the base score from an \emph{Exploitability} sub-score and an \emph{Impact} sub-score. Each metric value is first mapped to its numerical constant (e.g., $AV\in\{0.85,0.62,0.55,0.20\}$; $AC\in\{0.77,0.44\}$; $UI\in\{0.85,0.62\}$; and $C,I,A\in\{0,0.22,0.56\}$; while $PR$ depends on Scope). The equations are:
\begin{align}
\mathrm{Exploitability} &= 8.22 \times AV \times AC \times PR \times UI,\\
ISC_{\mathrm{Base}} &= 1 - (1-C)(1-I)(1-A),\\
\mathrm{Impact} &=
\begin{cases}
6.42 \cdot ISC_{\mathrm{Base}}, & S=U,\\
7.52\,(ISC_{\mathrm{Base}}-0.029) - 3.25\,(ISC_{\mathrm{Base}}-0.02)^{15}, & S=C,
\end{cases}\\
\mathrm{BaseScore} &=
\begin{cases}
0, & \mathrm{Impact}\le 0,\\
\mathrm{RoundUp}\!\left(\min(\mathrm{Impact}+\mathrm{Exploitability},\,10)\right), & S=U,\\
\mathrm{RoundUp}\!\left(\min(1.08(\mathrm{Impact}+\mathrm{Exploitability}),\,10)\right), & S=C.
\end{cases}
\end{align}
Here $\mathrm{RoundUp}(\cdot)$ rounds upwards to one decimal place (often written as $\mathrm{RoundUp}(x)=\lceil 10x\rceil/10$; implementations should avoid floating-point artifacts). The qualitative base severity rating is determined by the resulting base score using the standard bands: None (0.0), Low (0.1--3.9), Medium (4.0--6.9), High (7.0--8.9), and Critical (9.0--10.0).

\subsection{Case Study}

To illustrate why inferring CVSS metrics from an isolated vulnerable function, a single source file, or a commit diff can be misleading, we present a micro case study centered on \textbf{CVE-2020-21050}.

\begin{figure}[h]
\centering
\includegraphics[width=0.8\textwidth]{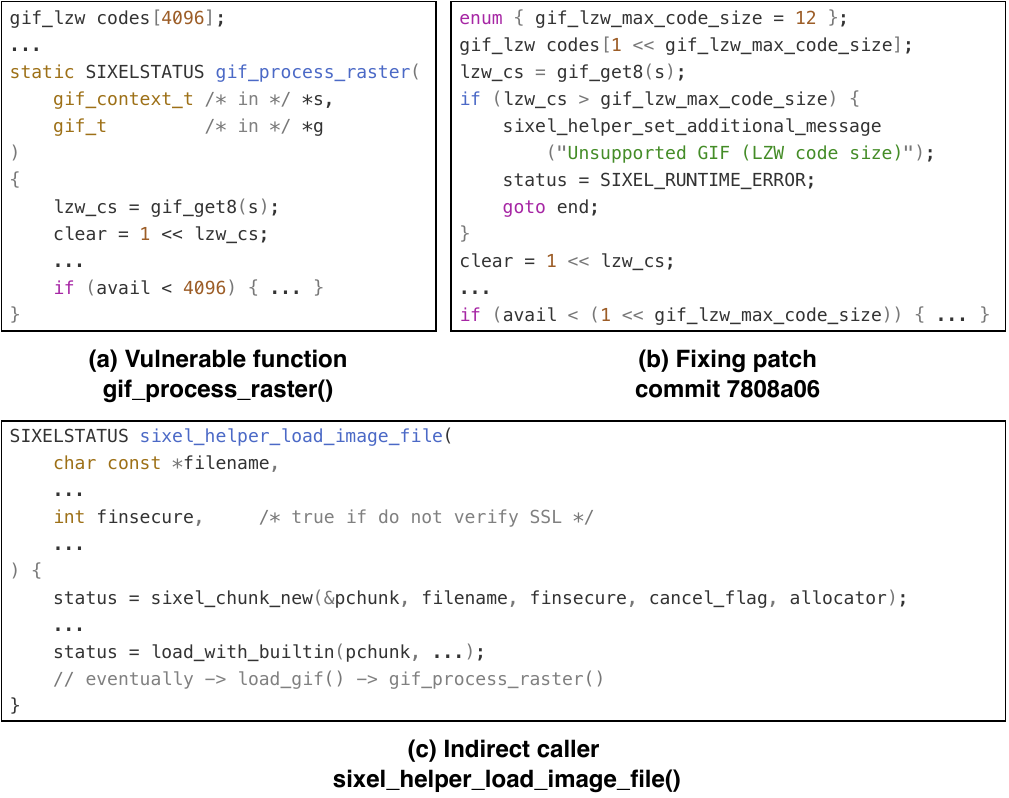}
\caption{Case study on assessing the Attack Vector (AV) of CVE-2020-21050. Neither the vulnerable function (a) nor the fix (b) contains sufficient evidence to infer AV. The indirect caller context (c) reveals the missing delivery semantics needed for AV assessment.}
\label{fig:case-study}
\end{figure}

According to the NVD record, Libsixel versions prior to v1.8.3 contain a stack buffer overflow in \texttt{gif\_process\_raster} in \texttt{fromgif.c} (Figure~\ref{fig:case-study}(a)). The upstream patch (Figure~\ref{fig:case-study}(b)) hardens the GIF LZW decoder by introducing a maximum supported LZW code size (12 bits), sizing the LZW code table based on this bound, and rejecting inputs whose LZW minimum code size exceeds it, thus preventing out-of-bound writes during decoding.

In CVSS v3.1, the Attack Vector metric captures how remotely an attacker can exploit the vulnerable component. \textit{Network (AV:N)} applies when the vulnerable component is tied to the network stack and exploitation is possible at the protocol level one or more network hops away, while \textit{Local (AV:L)} applies when the vulnerable component is not network-related and the attacker’s path is through local read, write, execute capabilities. In this case study, neither the body of \texttt{gif\_process\_raster} nor the fixing commit provides enough context to determine which delivery setting applies.

When considered in isolation, \texttt{gif\_process\_raster} is a delivery-agnostic parsing routine: it consumes bytes from an in-memory decoding state and performs LZW state transitions, without encoding whether those bytes originate from disk, stdin, or a network fetch. This can tempt an analyst to interpret the code as a local-only file parser and default to \textit{AV:L}. However, the NVD-published CVSS v3.1 vector for CVE-2020-21050 assigns \textit{AV:N}, motivating an examination of repository-level context to recover the missing delivery semantics.

The missing evidence emerges only when examining the indirect caller context. As shown in Figure~\ref{fig:case-study}(c), \texttt{sixel\_helper \_load\_image\_file} accepts an "insecure" flag (\texttt{finsecure}) that disables SSL certificate verification and delegates input materialization to a chunk-construction routine before dispatching the resulting byte buffer to format-specific decoders. This interface-level SSL control is consistent with builds and usages that support remote retrieval (e.g., via HTTPS in configurations that integrate \texttt{libcurl}). In such deployments, a malicious GIF hosted remotely can be fetched and decoded within the same component that receives the network data, supporting an \textit{AV:N} assessment.

This case study demonstrates that CVSS-relevant evidence may sit outside the vulnerable function, file, and patch context, and that omitting a small but critical portion of repository-level context can change a base-metric assignment (here, the difference between an \textit{AV:N} assessment based on repository-level information and an \textit{AV:L} assessment based on implementation-specific information).

\section{Methodology}

\label{sec:approach}

\subsection{Overview of \ourmethod}
\begin{figure}[h]
\centering
\includegraphics[width=0.9\linewidth]{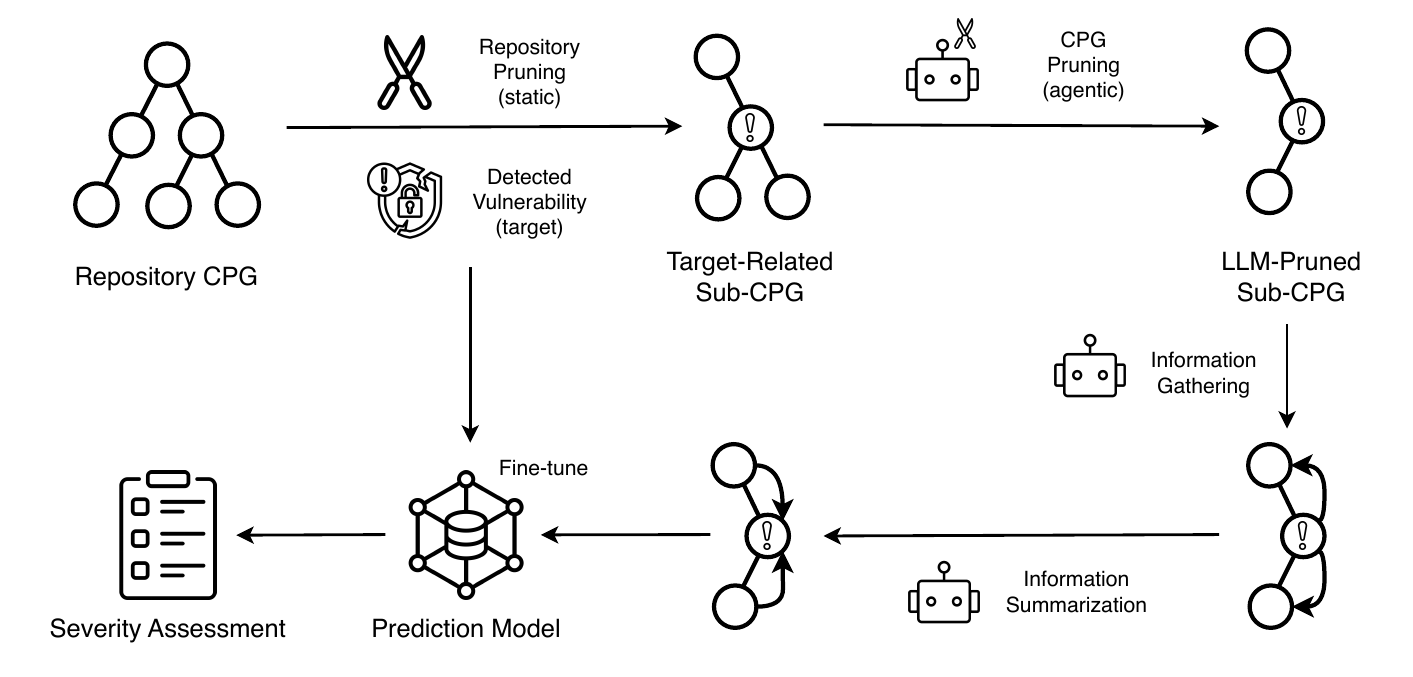}
\caption{Workflow of \ourmethod. Starting from the full repository CPG, \ourmethod first performs static repository pruning to obtain a target-related sub-CPG around the detected vulnerability. An agentic LLM then further prunes this subgraph into an LLM-Pruned sub-CPG. On top of the refined graph, \ourmethod iterates between information gathering that collects metric-relevant evidence from selected functions, and information summarization that aggregates per-metric implications. The resulting summaries are fed into a fine-tuned prediction model to output CVSS Base metrics and the overall severity assessment.}
\label{fig:workflow}
\end{figure}

Figure~\ref{fig:workflow} presents the workflow of \ourmethod, our repository-aware vulnerability severity assessment approach. When a vulnerability (hereafter “target”) is detected, \ourmethod first performs static repository pruning to remove unrelated regions from the full repository CPG. The pruned CPG and the target jointly define a target-related sub-CPG by traversing inter-procedural relationships to capture relevant context. Next, \ourmethod performs an agentic LLM-guided CPG pruning step that further refines this subgraph into an LLM-Pruned sub-CPG, focusing the context around CVSS-relevant behaviors. On top of the refined graph, \ourmethod runs an iterative loop with two stages: information gathering that selects metric-relevant evidence from the pruned graph and information summarization that consolidates the evidence into structured, per-metric implications for the CVSS Base metrics. Finally, these structured summaries are fed into a lightweight transformer-based prediction model, which is fine-tuned using task-aligned inputs and labels to predict the CVSS Base metrics. The derived severity assessment is then constructed based on these predicted metrics, capturing both the vulnerable code and its broader repository context.

\subsection{Static Repository Pruning}
\label{subsec:static-pruning}
The full repository CPG is usually too large to be processed directly by an LLM. Therefore, before invoking the agent, \ourmethod performs a lightweight static pruning step that keeps only functions that are structurally close to the vulnerable function.

We represent the repository call graph as a directed graph $G = (V, E)$, where each vertex $v \in V$ is a function, and each edge $(u,v) \in E$ indicates that $u$ may call $v$. Let $t \in V$ be the vulnerable (target) function. Because we care about both how the vulnerability is reached and how its effects propagate, we collect bounded neighborhoods in both directions. Using $D_{Caller}$ and $D_{Callee}$ as the maximum caller and callee depths, respectively, we form a candidate node set $V^{\text{cand}}$ as:
\begin{equation}
  V^{\text{cand}}
  = \{t\}
    \cup \mathrm{uniq}_{\phi}\!\Bigl(\mathcal{F}_{Caller}^{(D_{Caller})}(t)\Bigr)
    \cup \mathrm{uniq}_{\phi}\!\Bigl(\mathcal{F}_{Callee}^{(D_{Callee})}(t)\Bigr),
\end{equation}
where $\mathcal{F}_{Caller}^{(D_{Caller})}(t)$ contains all functions that can reach $t$ within $D_{Caller}$ caller hops, and $\mathcal{F}_{Callee}^{(D_{Callee})}(t)$ contains all functions reachable from $t$ within $D_{Callee}$ callee hops.
The operator $\mathrm{uniq}_{\phi}(\cdot)$ performs a lightweight de-duplication to suppress redundant nodes that contribute little additional structural coverage (e.g., repeated call sites or re-converging paths). Concretely, it keeps at most one representative per key $\phi(\cdot)$; in practice, $\phi$ can be chosen as a stable identifier such as (function name, defining file) to avoid conflating distinct local/static functions that share the same symbol. This de-duplication is applied to both caller and callee neighborhoods to further control neighborhood growth without relying on any learned filtering.

\subsection{Agentic LLM-Guided CPG Pruning and Bi-Directional Reasoning}
\label{subsec:agentic}
After static pruning, \ourmethod obtains a compact sub-CPG $G_0$ around the vulnerable function $t$. On top of this graph, we employ a large language model as an agent that explores caller and callee neighborhoods, prunes away irrelevant functions, and gathers per-metric evidence in a bi-directional manner. Pruning and information gathering happen simultaneously: whenever the agent decides to inspect a function, it both collects observations from that function and implicitly discards alternative branches.

\begin{figure}[h]
\centering
\includegraphics[width=0.7\linewidth]{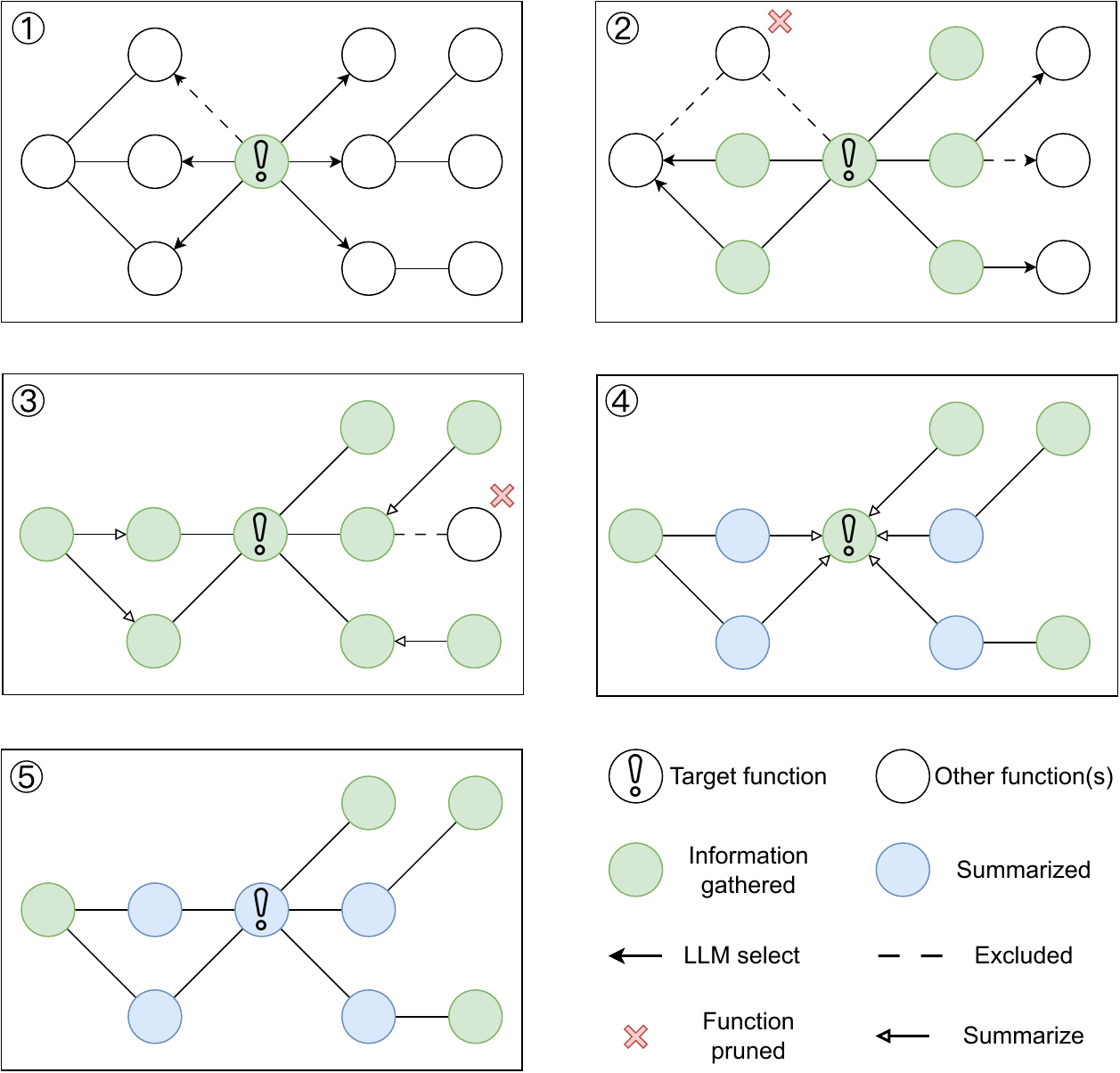}
\caption{Agentic LLM-guided pruning and bi-directional reasoning over a target-centered sub-CPG: \textcircled{1} initialization at the target, where the target function is inspected and marked as information gathered, while its neighbors remain undecided; \textcircled{2} the first gather-and-select step, where selected functions are marked as information gathered and unselected branches are excluded and pruned; \textcircled{3} continued function-wise exploration along selected caller/callee chains with additional gathered nodes; \textcircled{4} edge-to-center summarization starts, producing summaries for boundary functions first; \textcircled{5} summaries are propagated inward until the target receives the final per-metric summaries.}
\label{fig:llm-bidir}
\end{figure}

Figure~\ref{fig:llm-bidir} illustrates this process in five stages. Since \ourmethod explores both caller and callee neighborhoods symmetrically, absolute directional labels such as caller and callee are insufficient to express a function’s relative position with respect to the target. In particular, during exploration and pruning, what matters is not which side a function lies on, but how far it is from the target along the currently selected call chain, because our procedure expands outward to gather evidence and then aggregates that evidence back inward. Therefore, we introduce a target-centric notion of distance and use \textbf{center} and \textbf{edge} to denote relative proximity: functions closer to the target are \textbf{center functions}, whereas those farther away are \textbf{edge functions}. Importantly, these roles are relative and can change with the reference point—an intermediate function can be an edge function when viewed from the target, yet a center function when viewed from a more distant boundary node, capturing the inward–outward nature of our gather-and-summarize process.

\vspace{0.25em}
In Figure~\ref{fig:llm-bidir}, the left part of $G_0$ represents the caller-side context, which characterizes the preconditions and interaction pattern of the call and thus supports the exploitability-oriented metrics
\begin{equation}
  M_{Caller} = \{\mathrm{AC}, \mathrm{AV}, \mathrm{PR}, \mathrm{UI}, \mathrm{S}\},
\end{equation}
because these factors reveal how the vulnerability can be exploited, which are typically determined on the caller side.
In contrast, the right part represents the callee-side context, which describes the resources and behavior affected after the call executes and thus supports the impact-oriented metrics
\begin{equation}
  M_{Callee} = \{\mathrm{I}, \mathrm{C}, \mathrm{A}, \mathrm{S}\},
\end{equation}
since the actual loss in integrity, confidentiality, and availability is determined by the callee's state and operations.
As the metric scope ($\mathrm{S}$) can be influenced by both caller-side constraints and callee-side privilege changes, it appears in both $M_{Caller}$ and $M_{Callee}$.

\noindent\textbf{Stage \textcircled{1}: initialization at the target.}
At stage~\textcircled{1} in Figure~\ref{fig:llm-bidir}, the agent first focuses on the target function $t$.
The LLM is asked to produce an initial, observation-only description for each of the eight CVSS metrics as seen from $t$, and choose which callers and callees of $t$ should be analyzed next.
The result of this step includes per-metric observation lists associated with $t$, and a first set of selected neighbors that will be visited in the next round. The target is therefore already marked as "Information gathered", while surrounding functions are still undecided.

\noindent\textbf{Stages \textcircled{2}–\textcircled{3}: function-wise gathering and selection.}
The exploration phase then proceeds over the selected neighbors, as shown in stages~\textcircled{2} and~\textcircled{3}. At any moment, the agent maintains a queue of functions to visit on each side. For every function $f$ dequeued from this queue, the agent performs two actions in a single LLM call:
\begin{enumerate}
  \item Information Gathering: based on the full body of $f$ and the observations already collected so far (for $f$ itself and for its predecessors), the LLM updates the observation lists of the relevant metrics. LLMs are explicitly required to state observation from the code and from existing notes in short factual sentences and not to summarize or judge the value of the metrics.
  \item Select next functions: using the same context, the LLM chooses which callers and callees of $f$ should be analyzed next. These chosen neighbors are added to the queue; the remaining neighbors are treated as excluded for this analysis.
\end{enumerate}

This pattern of gather and select matches the transition from stage~\textcircled{1} to stage~\textcircled{2} and then to stage~\textcircled{3} in Figure~\ref{fig:llm-bidir}. Newly visited functions have their information gathered, while neighbors that are never selected are pruned from the CPG. The exploration phase stops once the queue is empty, which means that all functions deemed important by the agent have been visited at least once.

\vspace{0.25em}
\noindent\textbf{Stages \textcircled{4}–\textcircled{5}: edge-to-center summarization.}
After exploration, every selected function has a list of observation sentences for the metrics it contributes to, but these notes are still scattered across the graph. In the summarization phase, shown as stages~\textcircled{4} and \textcircled{5}, the agent compresses these per-function observations into short summaries that are anchored at the target.

Crucially, the summarization phase does not revisit the raw program graph, query additional function bodies, or re-use the observation of the current function. Instead, it operates on a stitched, injected representation constructed from the information gathered during exploration: caller-side contexts are placed before the current center function body, and callee-side contexts are inlined at the corresponding call sites. With the accumulated intermediate summaries, this injected view provides an execution-aligned context for compression, while keeping the evidence boundary fixed to what has already been collected.

The summarization proceeds from edge to center. Intuitively, the most peripheral functions do not receive information from any further functions, so there is no need to summarize them. The observations of these peripheral functions are then propagated inward. For a non-outermost function, the LLM sees the injected stitched view of the current function and is instructed to consolidate and compress it into a higher-level summary.

For metrics that are informed by a single direction (e.g., AV and UI on the caller side, or I and A on the callee side), the final metric text $S_m$ is exactly the summary obtained at the target in that direction. For Scope (S), which appears in both $M_{Caller}$ and $M_{Callee}$, we preserve both viewpoints by concatenating the caller-side and callee-side summaries. The collection of per-metric texts $\{S_m\}$ is then fed into the transformer-based predictor, which maps them to discrete CVSS Base metric scores. Because the predictor only consumes these observation-based summaries, the LLM's influence is limited to shaping the information presented to the small model rather than directly outputting CVSS decisions.

\begin{figure}[h]
\centering
\includegraphics[width=1.0\linewidth]{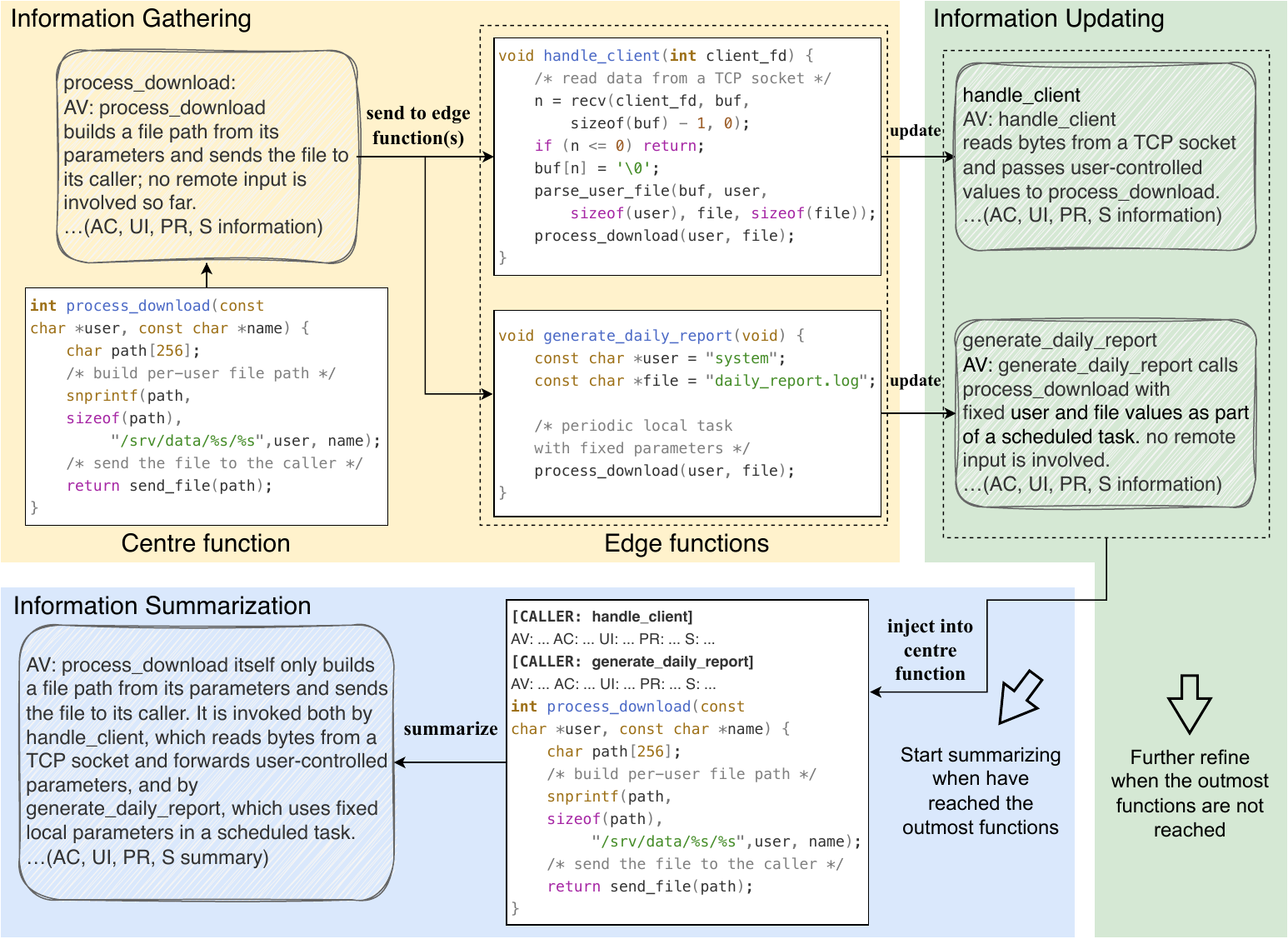}
\caption{Micro-case of evidence update, injection, and summarization for Attack Vector (AV). The top flow shows how AV-related observations are first gathered at the center function, then iteratively updated by exploring caller-side functions outward to the boundary. Before summarization, all collected evidence is injected into a stitched context. The summarizer then compresses this injected whole into a target-anchored AV explanation that is passed to the downstream predictor.}
\label{fig:micro-av-example}
\end{figure}

Figure~\ref{fig:micro-av-example} gives a concrete example of how \ourmethod collects, updates, and summarizes evidence for the Attack Vector (AV) metric. The figure groups the process into three blocks, which together illustrate how caller-side context is first accumulated and then consolidated into a target-anchored AV explanation.

In the Information Gathering block, analysis begins in the center function \texttt{process\_download} and produces an initial AV information record using only what is directly observable locally. At this point, the record can capture the usage of parameters and file-related operations, but it cannot determine the interaction channel by which these parameters are supplied. 

Next, selected caller-side functions are inspected in the Information Updating block. For each such function, the current AV information record associated with \texttt{process\_download} is reused as context, and observation-only sentences are updated based on the edge functions' code that reveals how inputs flow into the center function. Importantly, the update process is driven by the graph boundary: updating continues by moving outward along caller links until reaching the outmost functions of the explored subgraph. Only after the outmost functions have been reached, the process transitions from updating to summarization.

In the Information Summarization block, \ourmethod first injects the previously gathered and updated information into an integrated evidence context, and then summarizes based on it.
Concretely, caller-side contexts (including AV, AC, PR, UI, S) are placed before the center function body, while callee-side contexts (including I, C, A, S) are inserted at the corresponding call sites inside the center function, yielding a stitched, execution-relevant view of how the target is reached and what data flows into it. 
The summarizer then compresses this injected whole into a short metric-wise explanation anchored at the center function. This explanation consolidates what the explored callers reveal about how the center function becomes reachable, and provides the final metric-specific text passed to the downstream predictor. 

Overall, this micro-case highlights the intended workflow: CVSS metrics are not inferred from the target function alone, but are determined by progressively updating caller-side evidence, injecting all collected evidence into a unified context, and then summarizing back to a target-centered description.

Based on the evidence update and edge-to-center summarization processes described in Figures~\ref{fig:llm-bidir} and~\ref{fig:micro-av-example}, Algorithm~\ref{alg:agent-llm} formalizes the overall bi-directional reasoning procedure. Two LLM interfaces are used repeatedly. \textsc{LLMUpdateSelect} takes a function body and current evidence in one direction, appends observation-only sentences to the relevant metric evidence sets, and returns the next functions to inspect in that direction. \textsc{LLMSummarize} takes the injected representation of the current function together with already-produced child summaries, and produces per-metric summary texts that propagate from outmost functions back toward the target. In addition, \textsc{RevTopo} produces an edge-to-center processing order over the visited functions for a given direction.

\begin{algorithm}[t]
\caption{Agentic LLM-Guided Bi-Directional Reasoning}
\label{alg:agent-llm}
\begin{flushleft}
\textbf{Input:} Static subgraph $G_0$ with vulnerable function $t$;
caller metric set $M_{Caller}$; callee metric set $M_{Callee}$;\\
\textbf{Output:} Textual evidence summaries $S_m$ for all CVSS metrics $m$
\end{flushleft}
\begin{algorithmic}[1]
\For{each direction $d \in \{\text{Caller},\,\text{Callee}\}$}
  \State $\textit{ToVisit} \gets [\,t\,]$;\hspace{0.5em} $\textit{Visited} \gets \{t\}$;
  \State $E_m^{(d)}(f) \gets \emptyset$ for all functions $f$ and $m \in M_d$;
  \State $(\Delta E,\ \textit{Next}) \gets$ \Call{\textsc{LLMUpdateSelect}}{$t, d, M_d, G_0, \{E_m^{(d)}\}$};
  \State Update $E_m^{(d)}(t)$ with $\Delta E$;
  \State Enqueue all functions in \textit{Next} into \textit{ToVisit};
  \While{\textit{ToVisit} is not empty}
    \State Dequeue function $f$ from \textit{ToVisit};
    \State $(\Delta E,\ \textit{Next}) \gets$ \Call{\textsc{LLMUpdateSelect}}{$f, d, M_d, G_0, \{E_m^{(d)}\}$};
    \State Update $E_m^{(d)}(f)$ with $\Delta E$;
    \For{each selected function $g \in \textit{Next}$}
      \If{$g \notin \textit{Visited}$}
        \State Enqueue $g$ into \textit{ToVisit};
        \State Add $g$ to \textit{Visited};
      \EndIf
    \EndFor
  \EndWhile
  \State $\textit{Order} \gets$ \Call{\textsc{RevTopo}}{$G_0,\,\textit{Visited},\,d,\,t$};
  \For{each function $f$ in \textit{Order}}
    \State $\{S_m^{(d)}(f)\}_{m \in M_d} \gets$ \Call{\textsc{LLMSummarize}}{$f, d, M_d, \{E_m^{(d)}\}, \{S_m^{(d)}\}$};
  \EndFor
\EndFor
\For{each metric $m$}
  \If{$m \in M_{Caller} \cap M_{Callee}$}
    \State $S_m \gets S_m^{(\text{Caller})}(t)\ \Vert\ S_m^{(\text{Callee})}(t)$;
  \ElsIf{$m \in M_{Caller}$}
    \State $S_m \gets S_m^{(\text{Caller})}(t)$;
  \Else
    \State $S_m \gets S_m^{(\text{Callee})}(t)$;
  \EndIf
\EndFor
\State \Return $\{S_m\}$;
\end{algorithmic}
\end{algorithm}

The resulting $\{S_m\}$ provides a unified, target-anchored textual view of evidence collected from both directions, and is directly consumed by the downstream predictors.

\section{Experimental Setup}

\subsection{Dataset Construction}
\label{subsec:dataset}

We construct our dataset by starting from function-level vulnerability instances in the \textsc{PrimeVul}~\cite{ding2024vulnerability} and \textsc{TitanVul}~\cite{li2025out} benchmarks. While these benchmarks provide no CVSS annotations, they are curated to be high-quality and low-noise, and crucially, they link each vulnerability instance to a concrete repository state (commit hash) and a corresponding vulnerable function. Such curated linkage enables us to reliably localize the vulnerable function in a reproducible snapshot and recover the surrounding project context for downstream function-level parsing and static context extraction.

For each benchmark instance, we clone the referenced repository and check out the corresponding commit, and then perform function-level parsing and static context extraction following Section~\ref{subsec:static-pruning}. We discard instances that cannot be reproduced due to missing repositories, missing commits, checkout failures, or failures in parsing/context extraction. We then use the provided vulnerability identifier (e.g., CVE ID) to query NVD for the CVSS v3.x vector, treat the retrieved vector as supervision for downstream metric-wise prediction, and retain only instances with a complete CVSS v3.x vector available.

Some vulnerabilities span multiple modified hunks (or multiple files) within the same commit, which can result in near-duplicate function-level samples. To avoid over-counting and evaluation bias, we keep only one representative sample for each combination of CVE, repository, and commit. When there are multiple candidates, we randomly select one representative and remove the others. We additionally deduplicate overlapping instances across \textsc{PrimeVul} and \textsc{TitanVul} by matching CVE and function signature. When duplicates occur, we prefer \textsc{TitanVul} due to its more complete metadata and more reliable repository links.

To prevent data leakage, we split the dataset by CVE identifier such that no CVE appears in more than one of the training, validation, or test sets. We shuffle these CVEs and allocate them into train/validation/test splits with a ratio of 8:1:1, and additionally enforce that each split contains all possible values of each CVSS v3.x metric to mitigate missing-class issues in evaluation.
After filtering and de-duplication, we obtain 6,816 CVSS v3.x labeled metric instances, spanning 90 CWE types.

\subsection{Baselines}
\label{subsec:baselines}
We evaluate our approach against the following baselines.

\paragraph{Commercial and open-source LLM-based baselines.}
We consider LLM prompting baselines for CVSS v3.1 prediction under two widely used inference paradigms.
First, we adopt a direct prompting baseline, where the LLM is prompted to predict each CVSS v3.1 base metric from the vulnerability-related input, and the base severity is then derived from the resulting metric vector.
Second, motivated by previous work suggesting that agentic prompting can improve reasoning robustness and mitigate single-pass errors, we include a multi-agent baseline (EvalSVA-style~\cite{wen_evalsva_2024}) instantiated on the same LLM.
Concretely, we follow EvalSVA and organize agent interaction using the \emph{Preceding One Expert} communication strategy: each agent (except the first) conditions on the original input and the response from the immediately preceding expert, and the final prediction for each base metric is taken from the last expert.
We instantiate these two prompting strategies across both commercial and open-source LLMs to provide a broad and model-agnostic comparison.

\paragraph{Lightweight function-level models.}
Beyond commercial LLM baselines, we also compare against lightweight neural models that operate purely at the function level, which more closely resemble prior function-level vulnerability assessment methods.
Specifically, we include a series of Transformer-based models (including CodeBERT~\cite{feng_codebert_2020}, GraphCodeBERT~\cite{guo2020graphcodebert}, and UniXcoder~\cite{guo2022unixcoder}) that take the same vulnerability-related input as our main method but restrict evidence to function-level context.
These models are trained to predict all CVSS v3.1 base metrics and the resulting severity is derived accordingly.
This baseline highlights the incremental benefit brought by richer context information.

\subsection{Evaluation Metrics}
\label{subsec:eval-metrics}
We evaluate CVSS v3.x base severity as a multi-class classification task and report Accuracy, Macro-F1, and Weighted-F1 (abbreviated as Acc, M-F1, and W-F1 in the following paragraphs).
Macro-F1 averages F1 scores across labels and thus better reflects performance on minority classes, while Weighted-F1 weights each label by its support and reflects overall label consistency under class imbalance.
In addition, we computed the base severity derived from the predicted base metrics according to the CVSS v3.1 standard and evaluated it using the same set of metrics.

\subsection{Implementation Details}
\label{subsec:implementation}

For the main information collection and synthesis stage, we use GPT-4.1 and pin it to a fixed snapshot (\texttt{gpt-4.1- 2025-04-14}) to reduce potential drift caused by model updates.
To facilitate automatic parsing and downstream evaluation, we require every model to return output in a predefined JSON format. Concretely, our prompts explicitly specify the required JSON schema (fields, types, and constraints). We then perform strict JSON parsing and schema validation; if the output is malformed JSON or missing required fields, the system re-issues the request with corrective instructions to obtain a valid structured response.

All neural models are implemented in PyTorch using the HuggingFace \texttt{transformer} training framework and are trained on NVIDIA H100 GPUs. Unless otherwise specified, we fine-tune a shared encoder with eight task-specific linear classification heads to jointly predict the eight CVSS v3.1 base metrics, and optimize the mean cross-entropy loss over these heads. We use the AdamW optimizer with weight decay $0.01$~\cite{loshchilov2019decoupled}, an initial learning rate of $\eta = 2\times10^{-5}$, a linear learning rate schedule with a warmup ratio of $0.1$, and mixed-precision training (FP16). By default, the batch size per-device is $16$ with gradient accumulation steps of $2$, resulting in an effective batch size of $32$. We run experiments on three train/validation/test splits and report averaged results across splits.

\section{Evaluation Results}
Our experimental evaluation is designed to answer the following research questions (RQs):

\begin{itemize}
    \item \textbf{RQ1: How effectively does \ourmethod improve CVSS base severity prediction compared with LLM prompting baselines and function level predictors?}
    \item \textbf{RQ2: Which CVSS base metrics contribute most to the severity improvements brought by \ourmethod?}
    \item \textbf{RQ3: How consistent are the benefits of \ourmethod across different LLM instantiations and Transformer code encoder backbones?}
\end{itemize}

\subsection{RQ1: Overall effectiveness of \ourmethod on base severity prediction}
\label{subsec:rq1}

We evaluate the overall effectiveness of \ourmethod on base severity prediction by comparing it against prompting based LLM baselines, including commercial and open source models, and lightweight Transformer based code encoders trained on function level information.
All methods are assessed using Accuracy, Macro F1, and Weighted F1 on the derived CVSS v3.1 base severity label.

\begin{table*}[h]
\centering
\scriptsize
\caption{Base severity prediction performance comparison in \% across LLM prompting baselines and Transformer based function level predictors, together with \ourmethod.
Direct prompts the LLM once to predict the CVSS v3.1 base metric vector and derives base severity from the predicted vector.
EvalSVA follows the EvalSVA style multi agent prompting setup from prior work~\cite{wen_evalsva_2024} using the Preceding One Expert communication strategy and takes the last expert as the final prediction.}
\label{tab:severity-overall}
\resizebox{0.8\textwidth}{!}{%
\begin{tabular}{cccccc}
\toprule
Category & System / Model & Approach & Acc & M-F1 & W-F1 \\
\midrule

\multirow{4}{*}{Commercial LLM}
  & \multirow{2}{*}{GPT 4.1}
    & Direct  & 23.81 & 17.90 & 22.48 \\
  & & EvalSVA & 23.81 & 18.64 & 23.86 \\
  \cmidrule(lr){2-6}
  & \multirow{2}{*}{Claude Sonnet 4.0}
    & Direct  & 33.33 & 22.21 & 34.87 \\
  & & EvalSVA & 33.33 & 22.11 & 34.70 \\
\midrule

\multirow{7}{*}{Open source LLM}
  & \multirow{2}{*}{DeepSeek V3.2}
    & Direct  & 21.43 & 17.99 & 27.21 \\
  & & EvalSVA & 20.24 & 16.87 & 25.22 \\
  \cmidrule(lr){2-6}
  & \multirow{2}{*}{Qwen3 Coder 30B A3B}
    & Direct  & 27.38 & 17.31 & 28.34 \\
  & & EvalSVA & 34.52 & 23.86 & 36.65 \\
  \cmidrule(lr){2-6}
  & \multirow{2}{*}{Llama 3.1 8B}
    & Direct  & 17.86 & 11.72 & 18.42 \\
  & & EvalSVA & 15.48 & 12.00 & 19.16 \\
\midrule

\multirow{4}{*}{Transformer}
  & CodeBERT      & \textemdash & 35.32 & 18.81 & 27.30 \\
  \cmidrule(lr){2-6}
  & GraphCodeBERT & \textemdash & 37.30 & 24.08 & 36.67 \\
  \cmidrule(lr){2-6}
  & UniXcoder     & \textemdash & 41.27 & 29.62 & 41.96 \\
\midrule

\textbf{Proposed Method} & \textbf{\ourmethod} & \textemdash & \textbf{47.22} & \textbf{34.15} & \textbf{47.72} \\
\bottomrule
\end{tabular}%
}
\end{table*}

Table~\ref{tab:severity-overall} reports the base severity results.
Overall, \ourmethod achieves the best performance with 47.22\% accuracy, 34.15\% Macro F1, and 47.72\% Weighted F1, outperforming all LLM prompting baselines and all Transformer based function level predictors.
Compared to the best transformers-based function level UniXcoder model, \ourmethod improves accuracy from 41.27\% to 47.22\% and Macro F1 from 29.62\% to 34.15\%, indicating that the gains are not limited to majority class effects and that non dominant severity labels are better recognized.
The advantage of \ourmethod is also clear when compared against LLM based baselines.
The strongest LLM baseline is Qwen3 Coder 30B A3B with EvalSVA, yet it still trails \ourmethod by 36.8\% in accuracy and 43.1\% in Macro F1 relatively.
Even against the best commercial LLM result, Claude Sonnet 4 under Direct prompting, \ourmethod improves Macro F1 from 22.21\% to 34.15\% and Weighted F1 from 34.87\% to 47.72\%, showing that contextual augmentation yields substantially more reliable severity prediction than prompting alone.

We further observe that the EvalSVA-style multi-agent prompting strategy does not consistently improve model performance across different architectures.
For DeepSeek V3.2, EvalSVA leads to a slight degradation in performance, with accuracy decreasing from 21.43\% to 20.24\% and Macro F1 declining from 17.99\% to 16.87\%.
A similar trend is observed for commercial LLMs. Compared to Direct prompting, EvalSVA alters the final metric decisions in 11.90\% of cases for Claude Sonnet 4.0 and 1.34\% for GPT-4.1. However, these decision changes do not translate into substantial improvements in base severity prediction. Claude Sonnet 4 maintains the same accuracy (33.33\%), while its Macro F1 slightly decreases from 22.21\% to 22.11\%. GPT-4.1 exhibits only a marginal increase in Macro F1, from 17.90\% to 18.64\%, despite incurring more than twice the token consumption.

\begin{tcolorbox}[colback=gray!100,colframe=gray!400,title=Summary for RQ1]

\ourmethod achieves the best overall base severity prediction performance, consistently outperforming both Transformer-based function-level encoders and prompting-based LLM baselines. 
Compared with the strongest Transformer-based model, it improves accuracy and Macro F1 by 14.4\% and 15.3\% respectively. 
Against the strongest LLM baseline, the relative gains further increase to 36.8\% in accuracy and 43.1\% in Macro F1.

\end{tcolorbox}

\subsection{RQ2: Metric level effectiveness across CVSS base metrics}
\label{subsec:rq2}

To understand where the base severity improvements come from, we compare the strongest LLM prompting baseline, the strongest Transformer based function level encoder, and \ourmethod on each CVSS v3.1 base metric.
Specifically, we retain Qwen3 Coder 30B A3B with EvalSVA style multi agent prompting, UniXcoder trained on function level information only, and \ourmethod for this comparison.

\begin{table*}[h]
\centering
\scriptsize
\caption{Metric level performance comparison across CVSS v3.1 base metrics in \% for three representative systems.
LLM Prompt denotes the strongest prompting baseline in our setup, instantiated as Qwen3 Coder 30B A3B with EvalSVA style multi agent prompting where the final prediction is taken from the last expert.
Transformer denotes the strongest Transformer based function level predictor, instantiated as UniXcoder trained on function only evidence.
}
\label{tab:cvss-metric-breakdown}
\resizebox{\textwidth}{!}{%
\begin{tabular}{c*{4}{ccc}}
\toprule
Approach &
\multicolumn{3}{c}{Attack Vector} &
\multicolumn{3}{c}{Attack Complexity} &
\multicolumn{3}{c}{Privileges Required} &
\multicolumn{3}{c}{User Interaction} \\
\cmidrule(lr){2-4}\cmidrule(lr){5-7}\cmidrule(lr){8-10}\cmidrule(lr){11-13}
& Acc & M-F1 & W-F1
& Acc & M-F1 & W-F1
& Acc & M-F1 & W-F1
& Acc & M-F1 & W-F1 \\
\midrule
LLM Prompt &
63.10 & 31.68 & 61.23 &
75.00 & 42.86 & 82.65 &
72.62 & 43.05 & 69.44 &
72.62 & 54.51 & 67.49 \\
Transformer &
\textbf{73.41} & \textbf{41.39} & \textbf{72.08} &
92.86 & 48.14 & 89.44 &
\textbf{78.97} & 50.65 & \textbf{78.17} &
76.59 & 69.10 & 74.69 \\
\ourmethod &
67.86 & 38.53 & 66.28 &
\textbf{93.25} & \textbf{54.29} & \textbf{90.75} &
78.57 & \textbf{50.91} & 78.08 &
\textbf{79.37} & \textbf{75.29} & \textbf{79.23} \\
\bottomrule
\end{tabular}%
}

\vspace{1.0em}

\resizebox{\textwidth}{!}{%
\begin{tabular}{c*{4}{ccc}}
\toprule
Approach &
\multicolumn{3}{c}{Scope} &
\multicolumn{3}{c}{Integrity Impact} &
\multicolumn{3}{c}{Confidentiality Impact} &
\multicolumn{3}{c}{Availability Impact} \\
\cmidrule(lr){2-4}\cmidrule(lr){5-7}\cmidrule(lr){8-10}\cmidrule(lr){11-13}
& Acc & M-F1 & W-F1
& Acc & M-F1 & W-F1
& Acc & M-F1 & W-F1
& Acc & M-F1 & W-F1 \\
\midrule
LLM Prompt &
92.86 & 48.15 & 92.86 &
35.71 & 25.90 & 39.88 &
32.14 & 23.04 & 30.94 &
22.62 & 20.69 & 29.52 \\
Transformer &
\textbf{97.62} & \textbf{60.51} & \textbf{96.58} &
\textbf{55.56} & 39.68 & 54.36 &
\textbf{57.54} & 44.50 & 56.17 &
82.54 & 37.16 & 77.70 \\
\ourmethod &
\textbf{97.62} & \textbf{60.51} & \textbf{96.58} &
55.16 & \textbf{44.99} & \textbf{54.51} &
\textbf{57.54} & \textbf{45.51} & \textbf{56.66} &
\textbf{82.94} & \textbf{43.98} & \textbf{79.21} \\
\bottomrule
\end{tabular}%
}
\end{table*}

Table~\ref{tab:cvss-metric-breakdown} breaks down the performance by CVSS v3.1 base metrics and shows that the improvements of \ourmethod are concentrated on severity driving components.
Relative to the Transformer baseline, \ourmethod improves Macro F1 on Availability Impact from 37.16\% to 43.98\% and on Integrity Impact from 39.68\% to 44.99\%, while Confidentiality Impact increases from 44.50\% to 45.51\%.
On exploitation related metrics, \ourmethod shows clear gains on User Interaction, where Macro F1 increases from 69.10\% to 75.29\%, and on Attack Complexity, where Macro F1 increases from 48.14\% to 54.29\%.
Attack Vector remains less stable, with \ourmethod lower than the Transformer baseline in both accuracy and Macro F1, suggesting that Attack Vector often depends on cues that are not consistently strengthened by the injected context.

Compared with LLM Prompt, the advantage of \ourmethod is most pronounced on impact estimation.
For Integrity Impact, Confidentiality Impact, and Availability Impact, \ourmethod achieves Macro F1 values of 44.99\%, 45.51\%, and 43.98\%, while LLM Prompt attains 25.90\%, 23.04\%, and 20.69\% on the same metrics.
Scope shows limited headroom in this setting, since \ourmethod matches the Transformer baseline on Scope across all three metrics.

\begin{tcolorbox}[colback=gray!100,colframe=gray!400,title=Summary for RQ2]
\ourmethod delivers its strongest gains on consequence-oriented impact estimation, increasing Availability Impact Macro F1 from 37.16\% to 43.98\% and Integrity Impact Macro F1 from 39.68\% to 44.99\%.
It also improves exploitation precondition modeling, with User Interaction Macro F1 rising from 69.10\% to 75.29\% and Attack Complexity Macro F1 rising from 48.14\% to 54.29\%, indicating that injected context helps assemble dispersed cues into coherent metric-level decisions.
\end{tcolorbox}

\subsection{RQ3: Adaptability across Different LLM Instantiations and Transformer Backbones}
\label{subsec:rq3}

We further assess the adaptability of \ourmethod by varying two factors in the contextual augmentation pipeline.
The first factor is the final prediction backbone, where we consider CodeBERT, GraphCodeBERT, and UniXcoder.
The second factor is the LLMs used for contextual augmentation, where we instantiate the same pipeline with GPT-4.1 (\texttt{gpt-4.1-2025-04-14}), Claude Sonnet 4.0 (\texttt{claude-sonnet-4-20250514}), and DeepSeek-chat (the non-thinking mode of DeepSeek-V3.2).
For each backbone, we compare \ourmethod augmented variants against a model trained on the same function level data but without contextual augmentation, which serves as an ablation that removes the LLM driven contextual augmentation component while keeping the backbone fixed.
All configurations are evaluated on base severity using Accuracy, Macro-F1, and Weighted-F1.

\begin{table*}[h]
\centering
\scriptsize
\caption{Base severity performance across three function encoder backbones and three augmentation LLM instantiations.
For each backbone, the reported $\Delta$ values are the relative improvements over the corresponding Func-Level baseline trained on the same function only evidence.
DeepSeek-chat, Claude Sonnet 4.0, and GPT-4.1 are used only to instantiate the contextual augmentation stage, while the backbone is kept fixed within each block.}
\label{tab:severity-average-multiple-llms}
\resizebox{1.0\textwidth}{!}{%
\begin{tabular}{cccccccc}
\toprule
Backbone & Contextual Augmentation LLM
& Acc (\%) & $\Delta$Acc (\%)
& M-F1 (\%) & $\Delta$M-F1 (\%)
& W-F1 (\%) & $\Delta$W-F1 (\%) \\
\midrule

\multirow{3}{*}{CodeBERT}
& DeepSeek-chat
& 38.49 & +8.98
& 25.44 & \textbf{+35.26}
& 36.37 & \textbf{+33.22} \\
& Claude Sonnet 4.0
& 40.87 & \textbf{+15.72}
& 24.34 & +29.42
& 36.15 & +32.41 \\
& GPT-4.1
& 40.48 & +14.61
& 21.53 & +14.46
& 33.95 & +24.36 \\
\midrule

\multirow{3}{*}{GraphCodeBERT}
& DeepSeek-chat
& 39.29 & +5.32
& 26.02 & +8.05
& 36.86 & +0.52 \\
& Claude Sonnet 4.0
& 40.08 & +7.45
& 26.76 & \textbf{+11.12}
& 38.60 & \textbf{+5.25} \\
& GPT-4.1
& 42.86 & \textbf{+14.91}
& 26.35 & +9.43
& 37.95 & +3.49 \\
\midrule

\multirow{3}{*}{UniXcoder}
& DeepSeek-chat
& 44.84 & +8.65
& 31.98 & +7.97
& 44.41 & +5.84 \\
& Claude Sonnet 4.0
& 46.03 & +11.53
& 32.01 & +8.07
& 46.51 & +10.84 \\
& GPT-4.1
& 47.22 & \textbf{+14.42}
& 34.15 & \textbf{+15.29}
& 47.72 & \textbf{+13.73} \\
\bottomrule
\end{tabular}%
}
\end{table*}

Table~\ref{tab:severity-average-multiple-llms} shows that contextual augmentation improves base severity across all backbone and LLM combinations, as reflected by consistently positive relative improvements over the corresponding function level baseline.
The magnitude of improvement is backbone dependent.
On CodeBERT, the strongest improvements appear on Macro-F1 and Weighted-F1, reaching 35.26\% and 33.22\% relative improvement under DeepSeek-chat, suggesting that LLM driven contextual augmentation can substantially improve class balanced performance when the function encoder is limited.
On GraphCodeBERT, the Weighted-F1 gains remain modest, ranging from 0.52\% to 5.25\% despite accuracy gains up to 14.91\%, indicating that improvements can be metric specific for this backbone.
With UniXcoder, the gains are more uniform across metrics, and GPT-4.1 achieves the best overall results, reaching 47.22\% accuracy, 34.15\% Macro-F1, and 47.72\% Weighted-F1.

The choice of augmentation LLM also interacts with the backbone.
For CodeBERT, Claude Sonnet 4.0 yields the largest accuracy improvement of 15.72\%, while DeepSeek-chat yields the largest Macro-F1 and Weighted-F1 improvements of 35.26\% and 33.22\%.
For GraphCodeBERT, GPT-4.1 yields the largest accuracy improvement of 14.91\%, while Claude Sonnet 4.0 yields the largest gains on Macro-F1 and Weighted-F1 at 11.12\% and 5.25\%.
For UniXcoder, GPT-4.1 is consistently best across all three metrics, indicating that a stronger backbone can lead to more stable benefits from contextual augmentation.

\begin{tcolorbox}[colback=gray!100,colframe=gray!400,title=Summary for RQ3]
Across three backbones and three LLM instantiations, contextual augmentation consistently improves base severity.
The largest relative improvement is observed on CodeBERT, where Macro-F1 improves by 35.26\%, while the best absolute performance is achieved on UniXcoder with GPT-4.1, reaching 47.22\% accuracy and 34.15\% Macro-F1.
\end{tcolorbox}

\section{Related Work}

\label{sec:related}

Existing vulnerability severity assessment methods can be broadly organized into three streams: \textbf{Descriptions}, \textbf{Hybrid}, and \textbf{Code}. Across these streams, the trajectory is consistent: approaches evolve from shallow, single-input classifiers toward multi-input, LLM-enabled pipelines, gradually expanding from coarse severity prediction to metric-level assessments.

\textbf{Description-based approaches}. Early work established the feasibility of text-only prediction using engineered features and classical learners~\cite{khazaei_automatic_2016}. CNN-based models then improved severity estimation from CVE descriptions~\cite{han_learning_2017, nakagawa_character_2019}, while term-weighting schemes like TF-IGM enhanced lexical salience~\cite{kudjo_improving_2019}. Broader model sweeps further suggested that both neural and boosted-tree baselines can be effective when trained on CVE descriptions~\cite{sahin_conceptual_2019}. Moving beyond a single score, linear pipelines showed that per-metric prediction of the full base vector is tractable~\cite{elbaz_fighting_2020}; subsequent work added explicit extraction steps to better structure the input before TextCNN classification~\cite{sun_automatic_2023}. Transformer fine-tuning then became the dominant paradigm for full-vector inference, including BERT ensembles~\cite{shahid_cvss_bert_2021} and XLNet variants~\cite{shi_xlnet_cvss_2022}, while CNN baselines remained competitive for severity and exploitability prediction under clean splits~\cite{okutan_predicting_2022}. Two representative directions further enriched the textual channel: incorporating NVD-referenced pages as open-source intelligence to mitigate sparse CVE descriptions~\cite{kuehn_cvss_2023}; and applying curriculum schedules on issue reports to better handle label difficulty~\cite{pan_towards_practical_2024}. Most recently, studies explore using LLMs directly for vector inference, aiming for improved robustness under concept drift~\cite{manjunatha_cve_2024, marchiori_can_2025}.

\textbf{Hybrid-based approaches}. As purely textual signals can be underspecified, hybrid methods align modalities to reduce ambiguity and stabilize predictions. A pragmatic bridge combines CVE text with an existing CVSSv2 vector to estimate v3.x full vectors, effectively transferring prior metric structure across CVSS versions~\cite{nowak_machine_2021}. Text-code fusion then appears in several forms: sequence encoders for text with relational GATs over code graphs for severity prediction~\cite{dong_dekedver_2023, lai_recurrent_2015, rgat}; GraphCodeBERT with multi-task heads to couple modalities across severity metrics~\cite{du_bimodal_mtl_2024}; and prompt-tuned CodeT5 to reduce labeled-data needs while covering multiple severity dimensions~\cite{xue_prompt_tuning_sva_2025}. Beyond representation learning, demonstrations can also serve as hints for in-context LLM reasoning about severity~\cite{gao_sva_icl_2025, Ye-KG4VA-2025}. Finally, adding intention features (i.e., exploitability, scope, and impact) together with prompt tuning and MTL has been shown to stabilize severity estimates while enabling downstream guidance such as fix suggestions~\cite{shen2025vulstamp}. Overall, modality alignment (text, code, prior vectors/analysis) improves robustness to sparse descriptions and cross-domain drift.

\textbf{Code-based approaches}. In contrast to text-centric pipelines, code-first inputs like commits or functions can enable earlier estimation and provide stronger grounding in the change that introduced the vulnerability. Multi-task learning over commit diffs predicts the full base vector while encouraging cross-metric consistency~\cite{le_deepcva_2021}. Finer-grained slicing further centers the analysis on vulnerable functions, reducing noise from large code regions~\cite{le_fine_grained_2022}. Function-level studies benchmark both classical and deep models for vector prediction, highlighting a substantial training-time gap between traditional machine learning (ML) and deep learning (DL), while also showing that multi-task learning (MTL) helps DL models perform more effectively~\cite{nguyen_automated_2024}. Recently, multi-agent LLM judges operate on commit information to deliberate per-metric labels, explicitly optimizing for explainability and metric-wise consistency~\cite{wen_evalsva_2024}. Collectively, code-only methods trade some coverage of descriptive cues for earlier availability and more direct linkage to program changes.

\textbf{Takeaway}. Overall, most existing severity assessment methods still mainly rely on vulnerability descriptions; even when code is incorporated, the input is usually limited to isolated artifacts such as a function slice or a commit diff. In contrast, considering repository-level context provides a broader basis for reasoning about severity beyond single functions or commits, which motivates the repository-level context-aware severity assessment explored in this work.

\section{Threats to Validity}
\paragraph{Threats to External Validity.}
Our evaluation may be threatened by data leakage introduced by highly correlated instances. In practice, a single vulnerability can correspond to multiple functions, which often share substantial overlap in code patterns, data structures, and surrounding context \cite{ding2024vulnerability,li2024cleanvul,li2026beyond,li2026graph}. If such functions are simultaneously included in the dataset, samples from the same vulnerability may become near-duplicates across splits, allowing the model to benefit from shared artifacts instead of learning transferable cues. To reduce this risk, we retain only one representative function per vulnerability, rather than incorporating all associated functions, thereby limiting redundancy and improving the fidelity of the generalization assessment.

\paragraph{Threats to Internal Validity.}
A further threat concerns the skewed label distributions in several CVSS metrics. This imbalance reflects real-world reporting tendencies, but it can bias learning toward majority classes and inflate performance under metrics such as overall accuracy or micro-averaged scores. As a result, strong aggregate results may obscure poor behavior on minority classes that remain practically important. To provide a more informative assessment, we report macro-averaged F1, which weights classes equally and better captures performance on under-represented categories.

\paragraph{Control of Validity.}
We take two complementary steps to manage these concerns. First, we curb correlation-driven leakage by using a single representative function for each vulnerability, preventing closely related functions from appearing across splits. Second, we mitigate the impact of label skew in evaluation by emphasizing macro-averaged F1, which discourages majority-class dominance and yields a more balanced view of model behavior across classes.

\section{Conclusion and Future Work}
\label{sec:conclusion}

This paper studies automated CVSS v3.x severity assessment from the perspective of repository-level context. Many Base metrics depend on how a vulnerable routine is reached and how its effects propagate through the whole repository, making function- or file-centric evidence insufficient and prone to systematic misclassification. The core difficulty is therefore not merely accessing code, but reliably retrieving, filtering, and organizing dispersed repository evidence into metric-oriented signals that can support stable CVSS inference.

To address this challenge, we proposed \ourmethod, a context-aware severity assessment approach that operates directly on repository artifacts. \ourmethod constructs a code property graph from a detected vulnerable function and applies a two-stage pruning strategy to manage the scale and noise of real repositories. A lightweight static pruning step preserves structurally relevant neighborhoods, while an agentic LLM-guided pruning step further focuses the graph on CVSS-informative callers and callees and collects metrics-related evidence. The LLM consolidates the gathered information into compact, metric-wise summaries, enabling a lightweight transformer predictor to perform scalable inference without requiring long-context repository inputs at runtime.

Empirically, we release a higher-quality repository-level benchmark comprising 6,816 CVSS v3.x labeled metric instances spanning 90 CWE types, providing a more realistic testbed for future research. \ourmethod outperforms function-only baselines on severity classification. Compared with the best-performing function-level models, it improves accuracy by 14.4\% and macro-F1 by 15.3\%. Relative to the strongest direct-prompting baseline, \ourmethod achieves even larger gains, boosting accuracy by 36.8\% and macro-F1 by 43.1\%. These findings underscore the value of explicit repository-level evidence acquisition and metric-oriented organization for practical vulnerability management.

There are several directions remaining for future work. First, integrating additional repository artifacts (e.g., build configurations, deployment descriptors, and privilege boundaries expressed outside source code) may further strengthen evidence for metrics such as AV, PR, and S. Second, incorporating human-in-the-loop review could improve robustness when evidence is ambiguous or environment-dependent. Finally, extending the framework to newer scoring standards and broader vulnerability taxonomies may help unify automated severity assessment across evolving ecosystems.

\newpage

\bibliography{bibliography}
\bibliographystyle{ACM-Reference-Format}

\end{document}